\documentclass[aps,prb,twocolumn,superscriptaddress]{revtex4-1}

\usepackage{amsmath}
\usepackage{amsfonts}
\usepackage{amssymb}
\usepackage{color}
\usepackage{graphicx}
\usepackage{hyperref}
\usepackage{hyperref}
\usepackage{url}

\newcommand{\figwidth}{8.6cm}
 
\begin{document}

\title{Robust Edge States in Amorphous Gyromagnetic Photonic Lattices}

\author{Shampy Mansha}

\affiliation{Division of Physics and Applied Physics, School of Physical and Mathematical Sciences, \\
Nanyang Technological University, Singapore 637371, Singapore}

\author{Y.~D.~Chong}

\affiliation{Division of Physics and Applied Physics, School of Physical and Mathematical Sciences, \\
Nanyang Technological University, Singapore 637371, Singapore}

\affiliation{Centre for Disruptive Photonic Technologies, Nanyang Technological University, Singapore 637371, Singapore}

\email{yidong@ntu.edu.sg}

\begin{abstract}
We numerically study amorphous analogues of a two-dimensional photonic Chern insulator.  The amorphous lattices consist of gyromagnetic rods that break time-reversal symmetry, with the lattice sites generated by a close-packing algorithm.  The level of short-range order is adjustable, and there is no long-range order.  The topologically nontrivial gaps of the photonic Chern insulator are found to persist into the amorphous regime, so long as there is sufficient short-range order.  Strongly nonreciprocal robust  transmission occurs via edge states, which are shown to propagate ballistically despite the absence of long-range order, and to be exponentially localized along the lattice edge.  Interestingly, there is an enhancement of nonreciprocal transmission even at very low levels of short-range order, where there are no discernable spectral gaps.
\end{abstract}

\maketitle

\textit{Introduction}---Topologically nontrivial bandstructures have been the subject of intense research interest for nearly forty years, starting from the discovery of the quantum Hall effect \cite{vonKlitzing1980,MStone} and its link to topological band invariants \cite{TKNN,Simon1983,Haldane1988}.  The quantum Hall system is an example of a ``Chern insulator'', the simplest class of topological insulator \cite{Bernevig}.  The theoretical conditions for realizing a Chern insulator are (i) spatial periodicity (so a bandstructure can be defined) and (ii) broken time-reversal symmetry ($\mathcal{T}$) \cite{TKNN,Simon1983}.  Such conditions are not limited to condensed matter, but can also be achieved in suitably-designed photonic crystals \cite{Haldane2008,Raghu2008}.  In recent years, several photonic Chern insulators, featuring topologically-protected electromagnetic edge states, have been demonstrated \cite{Wang2008,Wang2009,Poo2011,Skirlo2015,Hafezi2011,Rechtsman2013,Hafezi2013}.  The field of ``topological photonics'' \cite{topological_photonics_review} has been fruitfully extended into many other areas, including anomalous Floquet insulators \cite{Hu2015,Gao2016,Leykam2016,Maczewsky2017,Mukherjee2017}, crystalline and valley-Hall insulators \cite{Wu2015,Barik2016,Ma2016,Dong2017}, and Weyl points \cite{Lu2013,Lu2015,Lin2016,Noh2017}.  Various groups have also explored the realization of topological band insulators using mechanical oscillators \cite{Nash2015,Wang2015}, acoustics \cite{Yang2015,CTChan2015,Fluery2016,Ni2017}, and electronics \cite{Jia2015}.

Due to the flexibility with which photonic structures can be designed and implemented, topological photonics provides new opportunities to study the interplay of disorder and band topology in two dimensions (2D), a topic of long-standing theoretical interest.  Topological band invariants, such as Chern numbers, are defined in periodic systems with Brillouin zones \cite{TKNN,Simon1983}; the quantum Hall case, where lattice periodicity is broken by a uniform magnetic field, can be handled by defining a magnetic Brillouin zone \cite{TKNN,Hofstadter} (a photonic analogue has been demonstrated by Hafezi \textit{et al.}~\cite{Hafezi2011,Hafezi2013}).  On the other hand, the introduction of weak disorder into a 2D $\mathcal{T}$-broken system is known to localize all or most of the Bloch states (although the localization lengths can be very large)~\cite{Laughlin1981,Halperin1982}.  Any extended states surviving in the bulk are percolating states that form thin bands embedded in the spectrum of localized states \cite{Chalker1988}.  The mobility gaps between these bands of extended states are spanned by robust edge states~\cite{Halperin1982}.  These features have been extensively investigated in quantum Hall systems \cite{Laughlin1981,Halperin1982,Chalker1988}, but Chern insulators of the ``quantum anomalous Hall'' type \cite{Haldane1988} should have the same universal behavior, even though their bands are not flat (i.e., not Landau level-like) in the ordered limit.

In this paper, we present a numerical study of 2D $\mathcal{T}$-broken photonic lattices that are \textit{amorphous}, possessing short-range positional order but no long-range order.  The lattices consist of microwave-scale gyromagnetic rods, with parameters very similar to previously-implemented gyromagnetic photonic crystals~\cite{Wang2009,Poo2011,Skirlo2015}, except that the sites are not placed periodically.  We investigate the behavior of the topological band gaps, and their edge states, as the level of short-range positional order is varied (by tuning the packing algorithm that generates the lattices).  In the limit of zero disorder, or perfect packing, the system is equivalent to a Chern insulator  of the ``quantum anomalous Hall'' type, with non-flat bands \cite{Haldane1988,Haldane2008,Raghu2008,Wang2008,Wang2009}.  In amorphous lattices, it is known that spectral gaps can continue to exist if there is sufficient short-range order.  Such gaps have been found, both experimentally and numerically, in $\mathcal{T}$-symmetric amorphous photonic structures~\cite{Ballato1999,Jin2001,Garcia2007,Florescu2009,HuiCao1,HuiCao2, CJin, Mansha2016}.  We show that for $\mathcal{T}$-broken photonic structures, topologically nontrivial higher-order gaps can likewise persist into the amorphous regime.  Despite the aforementioned theoretical arguments predicting that most 2D states must be localized \cite{Halperin1982}, we do not observe signatures of localization due to the limited system size.  We do, however, find evidence for robust edge states, in the form of strongly nonreciprocal transmission at frequencies coinciding with the ordered lattice's topological gaps.  Consistent with the concept of topological protection, these edge states can bypass sharp corners, and exhibit ballistic scaling of dwell time with propagation distance. Intriguingly, enhanced non-reciprocal transmission is observed even when the short-range order is so weak that the spectral gap cannot be discerned.  This appears to be consistent with the principle that robust edge states require only a mobility gap, not a spectral gap.  It is also reminiscent of earlier findings, in $\mathcal{T}$-symmetric amorphous structures, that even after a gap appears to close due to disorder, remnant effects such as Q-factor enhancements persist near the gap frequency~\cite{HuiCao1}.

Before proceeding, we take note of three recent related papers.  First, Mitchell \textit{et al.}~have shown that topological edge states can be realized in an irregularly-connected lattice of magnetically-interacting gyroscopes~\cite{Mitchell2016}, which can be regarded as an amorphous mechanical Chern insulator.  Second, Bandres \textit{et al.}~have shown numerically that robust edge states can exist in a $\mathcal{T}$-broken quasicrystal~\cite{Bandres2016}. Quasicrystals are lattices that are not periodic, but possess hidden long-range order; in the presence of $\mathcal{T}$-breaking, the spectrum is fractal and exhibits topological ``mini-gaps'' spanned by edge states~\cite{Bandres2016}, similar to the quasi-periodic Hofstadter model of the quantum Hall system \cite{Hofstadter}.  By contrast, the amorphous lattices studied here have only short-range order, and no long-range order.  Third, Agarwala and Shenoy \cite{Agarwala} have shown, with tight-binding models, that various classes of topological insulator behavior (including class A, or Chern insulators) can also occur on random lattices.

\begin{figure}  %----> Fig1
  \centering
  \includegraphics[width=\figwidth]{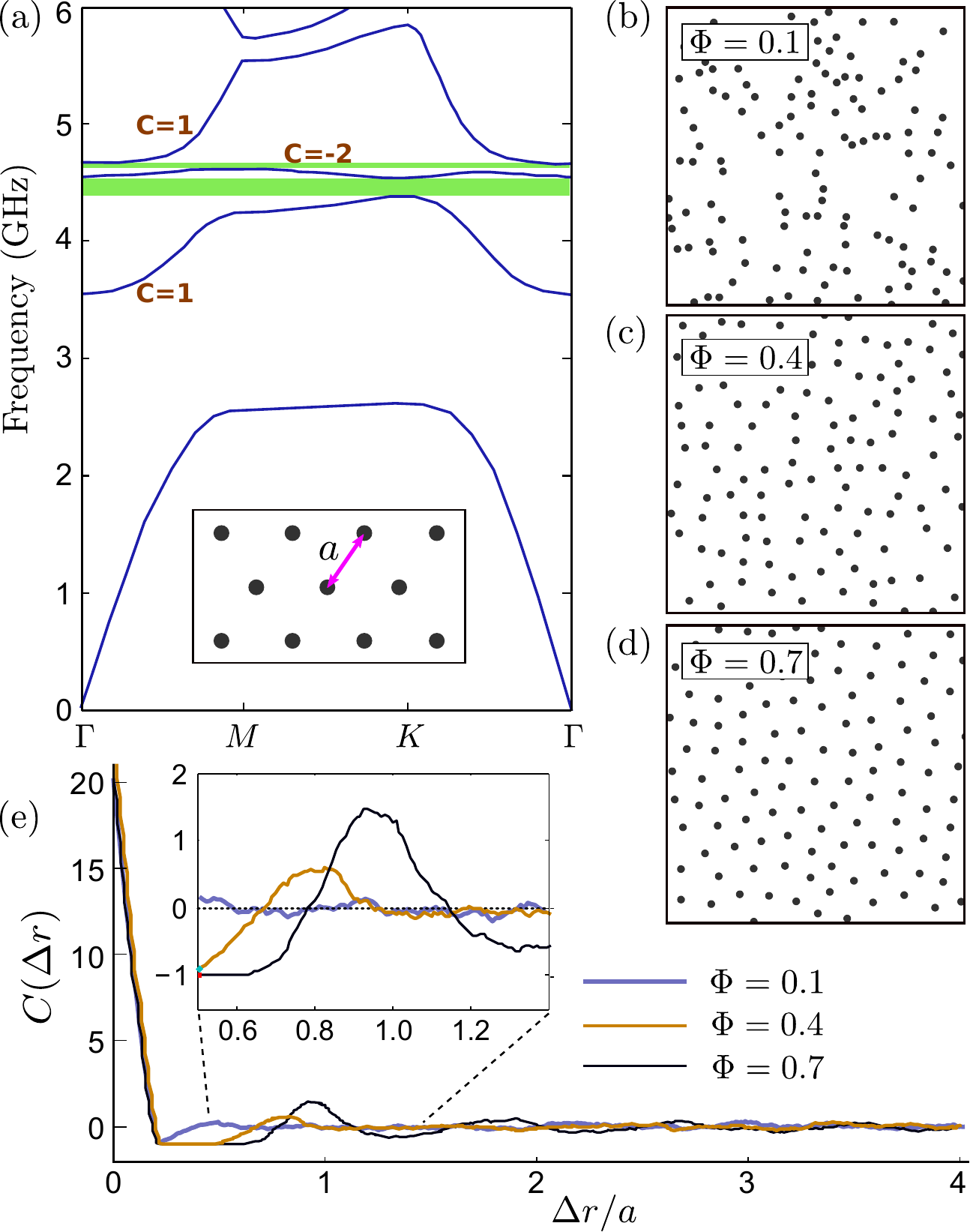}
  \caption{(a) Transverse magnetic (TM) band structure for a 2D triangular photonic crystal.  The Chern number $C$ for each topologically nontrivial band is indicated, and the topological band gaps are highlighted in green.  The structure (inset) consists of gyromagnetic rods with radius $r = 3.4\,\textrm{mm}$, nearest-neighbor spacing $a = 30.91\,\textrm{mm}$, permittivity $\epsilon=15$, and permeability components $\mu_{xx} = \mu_{yy} = 14$, $\mu_{zz}=1$, $\mu_{xy} = -\mu_{yx} = 10i$.  The rods are surrounded by air. (b)--(d) Amorphous lattices generated with packing parameters $\Phi=0.1$, $0.4$, and $0.7$. (e) Spatial correlation function $C(\Delta r)$ for different $\Phi$.  Each curve is averaged over 100 samples.  Inset: close-up of the nearest-neighbor correlation peak.}
  \label{fig:Bandstru_short_order}
\end{figure}

\textit{Lattice design}---We study 2D amorphous photonic lattices consisting of microwave-scale gyromagnetic rods made of a ferrite material (such as yttrium-iron-garnet), surrounded by air.  Similar structures have been used to experimentally realize photonic Chern insulators with topologically-protected edge states \cite{Wang2008,Wang2009,Poo2011,Skirlo2015}.  The system is sandwiched between metal plates, so that the fields are effectively 2D with only transverse magnetic (TM) modes: the electric field points out-of-plane (along $z$), and the magnetic field points in the $x$-$y$ plane.  Since the amorphous lattice sites are to be generated by a packing algorithm (see below), we first study the case of perfect packing, which corresponds to a triangular lattice.  We choose parameters so that the TM bandstructure of the triangular lattice, shown in Fig.~\ref{fig:Bandstru_short_order}(a), exhibits topological gaps.  Specifically, we take lattice constant $a = 30.91\,\textrm{mm}$ and rod radius $r = 0.11a$; the ferrite has dielectric constant $\varepsilon = 15$, and is magnetically biased in the $z$ direction to give permeability tensor components $\mu_{xx} = \mu_{yy} = 14$, $\mu_{zz}=1$, and $\mu_{xy} = -\mu_{yx} = 10i$.  These choices are close to previous experiments, except that those used square \cite{Wang2009,Skirlo2015} and honeycomb \cite{Poo2011} lattices.  The values of $r$, $\varepsilon$, and $\mu_{ij}$ are kept constant throughout the study.

Band gaps can be characterized by an integer $\Delta \mathcal{C}$, the sum of the Chern numbers for the bands below the gap.  According to the bulk-edge correspondence principle, $\Delta \mathcal{C}$ is equal to the net number of topological edge states in that gap \cite{Wang2008}.  In Fig.~\ref{fig:Bandstru_short_order}(a), we observe two topological band gaps, at frequencies $4.382\,\textrm{GHz} \le f \le 4.536\,\textrm{GHz}$ (with $\Delta \mathcal{C} = 1$) and $4.611\,\textrm{GHz} \le f \le 4.672\,\textrm{GHz}$ (with $\Delta \mathcal{C} = -1$).  In the non-gyromagnetic case $\mathbf{\mu} = \mathbf{1}$, the topological bandgaps close; the second and third TM bands meet at Dirac points at the $K$ and $K'$ points (the Brillouin zone corners), while the third and fourth bands have quadratic and Dirac band-crossing points at $\Gamma$ and along the Brillouin zone boundary.

Next, we generate amorphous lattices using molecular dynamics simulations \cite{Gao2006,HuiCao1}.  Each simulation contains 400 disks, bidisperse with radius ratio $1.2$ (50:50 distribution).  The disks are packed by steadily increasing their radii relative to the area in which they are enclosed, until a target filling fraction $\Phi$ is reached; the use of bidisperse disks allows larger $\Phi$ to be achieved before jamming.  The gyromagnetic rods are then placed at the disk centers (and the disks are then discarded).  A length scale is chosen so that for rod radius $r$, the rod filling fraction is the same as in the triangular photonic crystal of lattice constant $a$.  (Note that the final filling fraction of the rods is substantially less than the disk filling fraction $\Phi$.)  Finally, the lattice is truncated to the desired shape.

The amount of disorder is controlled by $\Phi$, which we call the ``packing parameter''.  As seen in Fig.~\ref{fig:Bandstru_short_order}(b)--(d), larger $\Phi$ produces more ordered lattices.  Even in the relatively ordered case shown in Fig.~\ref{fig:Bandstru_short_order}(d), corresponding to $\Phi = 0.7$, there is no long-range periodicity.  To quantify the positional order, we calculate the correlation function
\begin{equation}
  C(\Delta r)\equiv \langle \Theta(\mathbf{r}+ \Delta\mathbf{r})  \Theta(\mathbf{r})\rangle/\langle \Theta(\mathbf{r})\rangle^{2} - 1,  
\end{equation}
where $\Theta=1$ inside the rods and $\Theta=0$ outside.  The results are shown in Fig.~\ref{fig:Bandstru_short_order}(e) for several values of $\Phi$.  Each curve is averaged over 100 independent samples, and over 450 spaced-out choices of $\mathbf{r}$ for each sample.  The correlation decreases to zero for large $\Delta r$, corresponding to the absence of long-range order.  For all but the most disordered ($\Phi = 0.1$) samples, we observe a first-order correlation peak at $\Delta r \approx a$, indicating the typical nearest-neighbor spacing.  The prominence of this peak is a measure of the level of short-range order, and we indeed observe that it is stronger for larger values of $\Phi$.

\begin{figure}
  \centering
  \includegraphics[width=\figwidth]{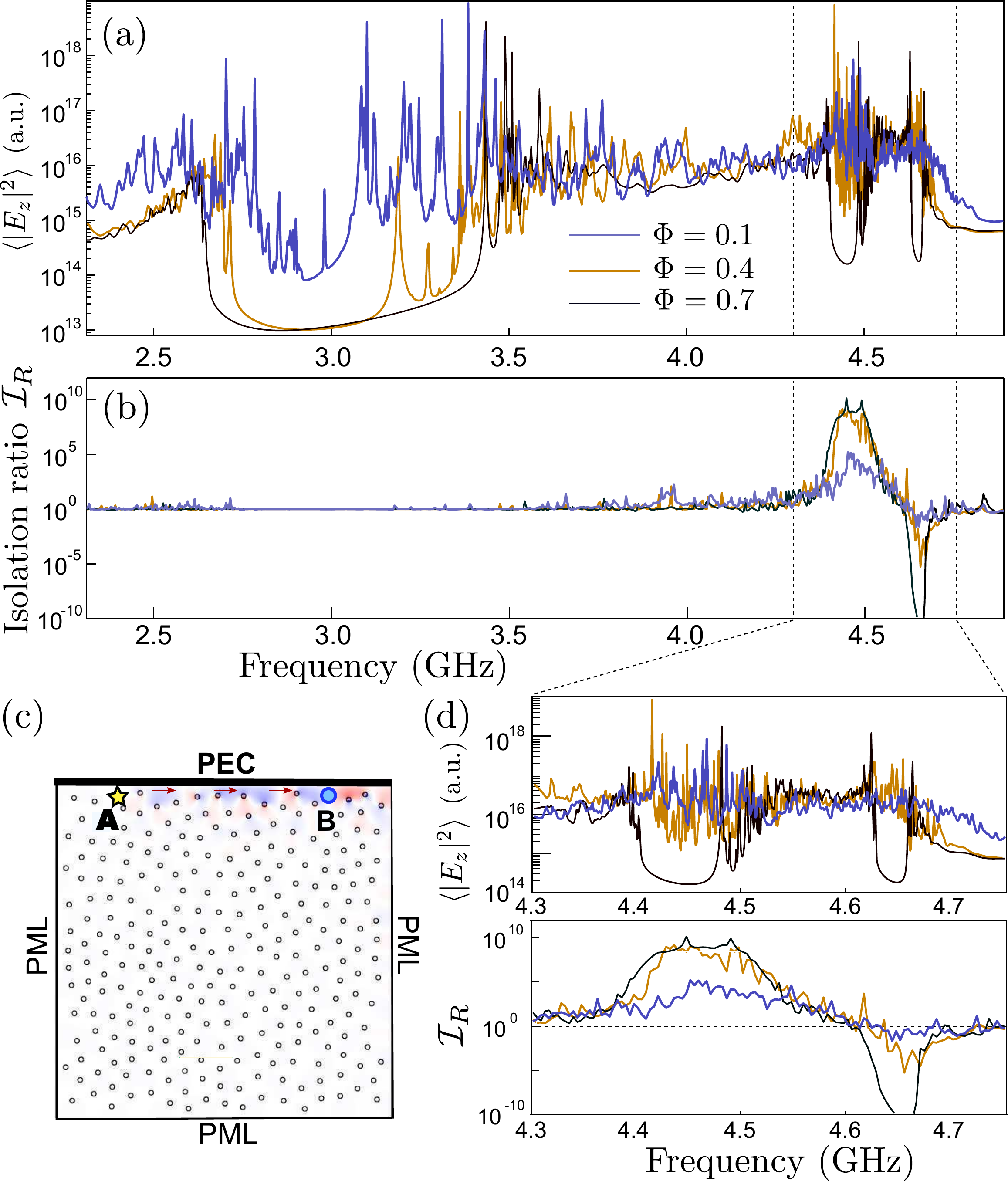}
  \caption{(a) Spectral density of amorphous lattices with different values of the packing parameter $\Phi$.  These results are obtained from frequency-domain calculations for an amorphous lattice of 266 rods distributed within a square region, surrounded by empty space, with a point transverse-magnetic (TM) source of fixed current amplitude at the center.  Each curve shows $|E_z|^2$ averaged over the lattice region, and over 10 samples.  For $\Phi = 0.7$, three spectral dips are observed, which diminish with greater disorder (smaller $\Phi$).  (b) Isolation ratio $\mathcal{I}_R=I_{AB}/I_{BA}$ versus frequency, for transmission along the lattice edge.  Each curve is averaged over 10 samples.  There is no enhancement of $\mathcal{I}_R$ at the primary gap, but the two higher-order gaps show enhancemed $\mathcal{I}_R$ corresponding to forward and backward transmission respectively.  (c) Schematic for the calculation in (b).  The upper edge is a perfect electrical conductor (PEC), while perfectly-matched layer (PML) boundary conditions are applied on the other three sides.  The isolation ratio is calculated between the two points labeled $A$ and $B$, near the edge.  Also shown is the steady-state real electric field $E_z$ emitted by a $4.448\,\textrm{GHz}$ point source at $A$, with red and blue corresponding to positive and negative values of $E_z$; the lattice packing parameter is $\Phi = 0.7$.  (d) Close-ups of (a) and (b) in the vicinity of the isolation ratio peaks.}
  \label{fig:DOS_Isolationratio}
\end{figure}

\textit{Bulk and edge states}---Fig.~\ref{fig:DOS_Isolationratio}(a) shows an estimate of the electromagnetic spectral weight of the amorphous lattices, obtained from finite-element (COMSOL Multiphysics) calculations of steady-state solutions for a point TM source in a finite lattice surrounded by empty space.  The field intensity $|E_z|^2$ is averaged over the lattice region, and then over 10 independent lattice realizations.  For $\Phi = 0.7$, we observe three prominent dips, coinciding with the triangular lattice band gaps shown in Fig.~\ref{fig:Bandstru_short_order}(a).  With increasing disorder (smaller $\Phi$), the dips become less prominent, confirming that the band gaps are induced by short-range order.  This is consistent with previous studies of $\mathcal{T}$-symmetric amorphous photonic lattices, which found photonic density-of-states dips conciding with topologically trivial gaps of an ``equivalent'' crystalline lattice \cite{HuiCao1,HuiCao2}.  Our results demonstrate that the same is true of topologically nontrivial gaps.

In order to determine whether the two higher-order dips in Fig.~\ref{fig:DOS_Isolationratio}(a) are associated with edge states, we pick two points $A$ and $B$ lying near a perfect electrical conductor (PEC) boundary, and compute the isolation ratio $\mathcal{I}_R = I_{AB}/I_{BA}$, where $I_{AB}$ is the field intensity $|E_z|^2$ at $B$ produced by a constant-frequency point $E_z$ source at $A$, and vice versa for $I_{BA}$.  The results are shown in Fig.~\ref{fig:DOS_Isolationratio}(b), where each curve is averaged over 10 independent samples of the same $\Phi$.  The simulation setup is shown in Fig.~\ref{fig:DOS_Isolationratio}(c).  Perfectly-matched layers (PMLs) are placed along the other three sides of the computational domain to completely absorb impinging waves.

\begin{figure}  %----> Fig3
  \centering
  \includegraphics[width=\figwidth]{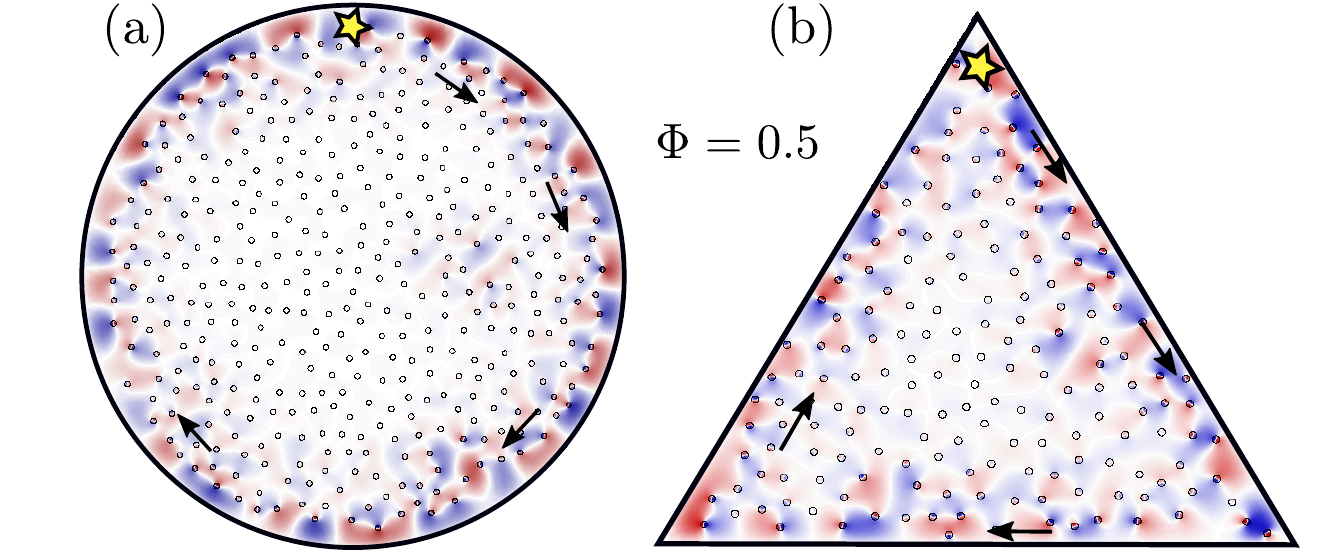}
  \caption{Robust TM edge propagation in amorphous lattices of $\Phi = 0.5$ with (a) circular and (b) triangular shape, and PEC along all edges.  The steady-state real electric field $E_z$ is indicated in red (positive) and blue (negative).  Stars indicate point sources emitting at $4.437\,\textrm{GHz}$.}
  \label{fig:Ez_plot}
\end{figure} 

Since $\mathcal{T}$ is broken by the gyromagnetic medium, electromagnetic wave propagation is non-reciprocal \cite{Jalas2013}, and $\mathcal{I}_R$ can deviate from unity at any frequency.  In Fig.~\ref{fig:DOS_Isolationratio}(b), however, we observe that the transmission is only strongly non-reciprocal (large $|\mathcal{I}_R|$) in two frequency ranges, which correspond to the second and third spectral dips in Fig.~\ref{fig:DOS_Isolationratio}(a).  The sign of $\mathcal{I}_R$ is different in the two ranges, consistent with the opposite signs of $\Delta \mathcal{C}$ in the two gaps.  This shows that short-range order can induce robust unidirectional transmission, with characteristics highly similar to the topological edge states of the gyromagnetic photonic crystal.  Interestingly, in Fig.~\ref{fig:DOS_Isolationratio}(b) and (d) an enhancement of $|\mathcal{I}_R|$ is observable in those frequencies ranges even for the $\Phi = 0.1$ case, where the disorder is so strong that the correlation function has no nearest-neighbor peak [Fig.~\ref{fig:Bandstru_short_order}(e)] and there is no clear spectral dip [Fig.~\ref{fig:DOS_Isolationratio}(a)].  Note also that $\mathcal{I}_R \approx 1$ throughout the primary gap, which is topologically trivial in origin.

The spatial distribution of the $E_z$ field is also shown in Fig.~\ref{fig:DOS_Isolationratio}(c), confirming that propagation occurs mainly along the edge.  Fig.~\ref{fig:Ez_plot} shows field distributions for circular and triangular boundaries, demonstrating propagation around curves and sharp corners.

\begin{figure}  %----> Fig4
  \centering
  \includegraphics[width=\figwidth]{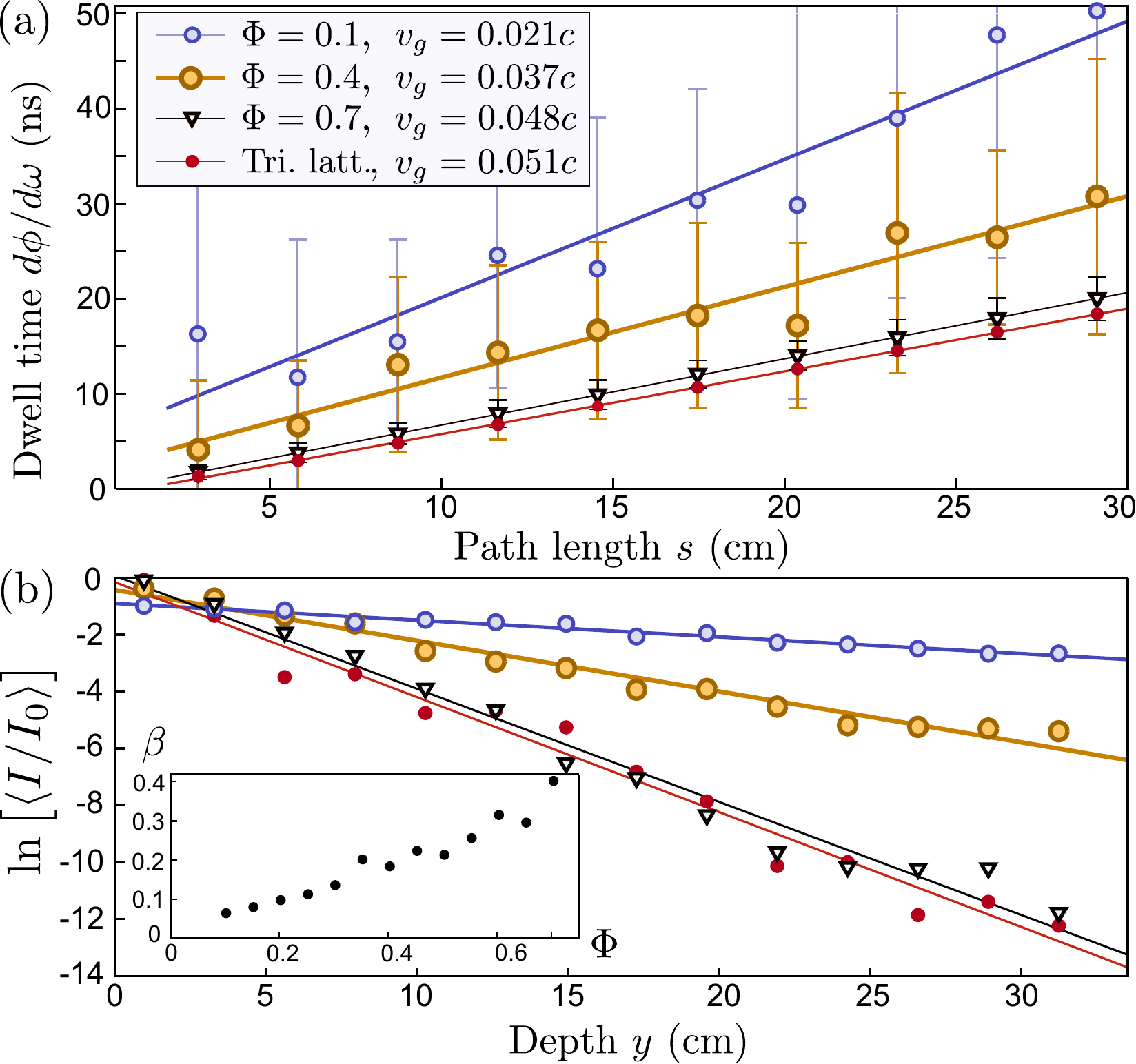}
  \caption{(a) Dwell time $d\varphi / d\omega$ versus path length $s$, for amorphous lattices with different packing parameters $\Phi$ as well as the perfectly-ordered triangular lattice.  Each symbol is averaged over 10 samples; error bars show the standard deviation.  The calculation procedure is explained in the text.  The legend indicates fitted values of the group velocity $v_g$, where $c$ is the speed of light.  (b) Semi-logarithmic plot of the intensity $I$ versus the distance from the edge $y$, where $I$ is the value of $|E_z|^2$ averaged over positions to the right of a point source, and over a frequency band and 10 lattice samples.  The results show that the waves are exponentially localized to the edge, $I \propto \exp(-\beta y)$.  Inset: plot of the fitted values of $\beta$ versus $\Phi$ (including $\Phi$ values omitted from the main plot for clarity).}
  \label{fig:Groupvelocity_intensity}
\end{figure}

A more rigorous method for demonstrating the robustness of edge propagation is to show that it is ballistic under disorder.  Ballistic topological edge transport has previously been established in a perfectly-ordered bulk lattice with a deformed edge \cite{Khanikev}.  To discover whether edge propagation in the amorphous lattices is similarly ballistic, we calculate the dwell time $d\varphi/d\omega$ as a function of propagation distance $s$ along a straight edge.  The simulation setup is similar to Fig.~\ref{fig:DOS_Isolationratio}(c), except that there are several point probes placed at equally-spaced intervals near the edge.  Each probe measures the local phase $\varphi$ over 21 equally-spaced frequencies in $4.4292\,\textrm{GHz} \le f \le 4.4677\,\textrm{GHz}$; this range is chosen to lie within the ordered lattice's first topological gap, and coincides with the peak in $\mathcal{I}_R$ shown in Fig.~\ref{fig:DOS_Isolationratio}(a).  For each lattice realization, $\varphi$ is found to have a linear relationship with $\omega$, with coefficient of determination $R^2 > 0.9$ for all tested samples in the range $0.1\le \Phi \le 0.7$.  We then calculate $d\varphi/d\omega$ (where $\omega = 2\pi f$) by a linear least-squares fit for each probe.  The procedure is repeated for 10 independent lattice realizations to obtain ensemble means.  The results are shown in Fig.~\ref{fig:Groupvelocity_intensity}(a).  For each value of $\Phi$, the mean dwell time appears to scales linearly with $s$, indicating ballistic behavior.  The slope is the inverse of the ensemble-averaged group velocity $v_g$.  For large $\Phi$, $v_g$ is very close to the value in the triangular lattice, but it is lower in more disordered lattices.

Finally, we verify in Fig.~\ref{fig:Groupvelocity_intensity}(b) that the propagating waves are exponentially localized to the edge.  The setup is similar to Fig.~\ref{fig:DOS_Isolationratio}(c), but with probes placed in a grid (with $x$-spacing $1.1a$ and $y$-spacing $0.75a$, and 11 probes for each $y$).  The normalized intensity $I = |E_z|^2$ is averaged over the probes, over 21 equally-spaced frequencies in $4.4292\,\textrm{GHz} \le f \le 4.4677\,\textrm{GHz}$, and over 10 independent lattice realizations for each $\Phi$.  Using these results, we perform a fit $I \propto \exp(-\beta y)$ to extract the inverse penetration depth $\beta$, which is found to increase with $\Phi$ (i.e., more ordered lattices exhibit stronger confinement).

\textit{Conclusions}---We studied amorphous lattices generated using a packing algorithm, with controllable levels of short-range order~\cite{Gao2006,HuiCao1}.  For strong short-range order, the lattices exhibit spectral dips that are obvious counterparts of the ordered lattice's photonic band gaps---not just the topologically trivial primary gap, but also the topologically nontrivial higher-order gaps.  Strongly non-reciprocal, ballistic, edge-localized transmission is observed at frequencies corresponding to the topologically nontrivial gaps.  This confirms the principle that topological edge states should be protected against disorder; in the photonic context, this protection had previously been probed using localized defects or edge deformations overlaid on an ordered bulk~\cite{Wang2008,Wang2009,Khanikev}.  Moreover, edge state dominated transmission is observed even when the short-range order is so weak that the spectral dips are not discernable, which appears to be consistent with theoretical arguments that edge states in disordered systems rely on the existence of a mobility gap, not a spectral gap \cite{Halperin1982}.  The evidence is not conclusive, however, because our sample sizes are too small to distinguish between localized and extended bulk states (and hence to locate the mobility gaps).  Larger-scale numerical studies might be able to identify the percolating bulk states, and probe their relationship with the edge states.

The lattice parameters that we have chosen are consistent with previously-reported experiments~\cite{Wang2009,Poo2011,Skirlo2015}.  Hence, these amorphous photonic lattices may be a fruitful experimental photonic platform for studying various effects involving topological band structures and disorder, such as topological Anderson insulators \cite{Li2009,Titum2015} and the floating or annihilation of extended states \cite{Khmelnitskii1984,Laughlin1984,Kivelson1992,Liu1996}.

We are grateful to J.~C.~W.~Song, G.~Vignale, S.~Adam, and E.~Berg for helpful discussions.  This research was supported by the Singapore MOE Academic Research Fund Tier 2 (grant MOE2015-T2-2-008) and the Singapore MOE Academic Research Fund Tier 3 grant MOE2011-T3-1-005.

%% Floating:
%% D.E. Khmelnitskii, Phys.Lett. A106, 182 (1984).
%% R.B. Laughlin, Phys.Rev.Lett. 52, 2304 (1984).

\end{document}